\documentstyle[11pt]{article}
\newcommand{\nc}{\newcommand}
\nc{\vare}{\varepsilon} \nc{\oi}{{0i}} \nc{\Mc}{{\mu}}
\nc{\psib}{\overline{\psi}} \nc{\psid}{{\psi^{\dagger}}}
\nc{\bb}{\begin{equation}} \nc{\ee}{\end{equation}}
\nc{\qq}{\qquad\qquad} \nc{\erm}{{\rm e}} \nc{\munu}{{\mu\nu}}
\nc{\dis}{\displaystyle} \nc{\um}{{1\over 2}}
\nc{\R}{{\rm I\!\!R}} \nc{\C}{{\rm I\!\!\!C}}
\nc{\vecna}{\mbox{\boldmath $\nabla$}}
\nc{\pa}{\partial} \nc{\ug}{\; = \;} \nc{\vs}{\vspace*}
\nc{\Hc}{{\cal H}}  \nc{\Lc}{{\cal L}}  \nc{\Lcn}{{\cal L}^{(n)}}
\nc{\Lcuno}{{\cal L}^{(1)}}  \nc{\Lcdue}{{\cal L}^{(2)}}
\nc{\Lctre}{{\cal L}^{(3)}}  \nc{\Lczero}{{\cal L}^{(0)}}
\nc{\Ebf}{\mbox{\boldmath $E$}} \nc{\Hbf}{\mbox{\boldmath $H$}}
\nc{\Vbf}{\mbox{\boldmath $V$}} \nc{\Fbf}{\mbox{\boldmath $F$}}
\nc{\Wbf}{\mbox{\boldmath $W$}} \nc{\lbf}{\mbox{\boldmath $l$}}
\nc{\xbf}{\mbox{\boldmath $x$}} \nc{\ubf}{\mbox{\boldmath $u$}}
\nc{\odbf}{\overline{\mbox{\boldmath $d$}}}
\nc{\okbf}{\overline{\mbox{\boldmath $k$}}}
\nc{\vbf}{\mbox{\boldmath $v$}} \nc{\wbf}{\mbox{\boldmath $w$}}
\nc{\jbf}{\mbox{\boldmath $j$}} \nc{\mubf}{\mbox{\boldmath $\mu$}}
\nc{\sigbf}{\mbox{\boldmath $\sigma$}}
\nc{\abf}{\mbox{\boldmath $a$}} \nc{\bbf}{\mbox{\boldmath $b$}}
\nc{\sbf}{\mbox{\boldmath $s$}} \nc{\dbf}{\mbox{\boldmath $d$}}
\nc{\rbf}{\mbox{\boldmath $r$}} \nc{\kbf}{\mbox{\boldmath $k$}}
\nc{\Lbf}{\mbox{\boldmath $L$}} \nc{\imp}{\mbox{\boldmath $p$}}
\nc{\albf}{\mbox{\boldmath $\alpha$}}
\nc{\dddov}{{\stackrel{\ldots}{v}}} \nc{\dox}{\dot{x}} \nc{\ddox}{\ddot{x}}
\nc{\dddox}{{\stackrel{\ldots}{x}}} \nc{\dopi}{\dot{\pi}} \nc{\dop}{\dot{p}}
\nc{\dov}{\dot{v}} \nc{\ddov}{\ddot{v}} \nc{\doa}{\dot{a}} \nc{\ddoa}{\ddot{a}}
\nc{\dddoa}{{\stackrel{\ldots}{a}}} \nc{\ddddoa}{{\stackrel{....}{a}}}
\nc{\ddddov}{{\stackrel{....}{v}}}
\nc{\Omn}{\Omega^\munu} \nc{\po}{\widehat{p}}
\nc{\ga}{\gamma} \nc{\al}{\alpha} \nc{\gm}{{\ga^\mu}} \nc{\gn}{{\ga^\nu}}
\nc{\gao}{\gamma^0} \nc{\gabf}{\mbox{\boldmath $\gamma$}}
\nc{\ao}{\widehat{a}} \nc{\vo}{\widehat{v}}
\nc{\So}{\widehat{S}} \nc{\Go}{\widehat{G}}
\nc{\rd}{{\rm d}} \nc{\dtau}{{\rd\tau}} \nc{\dt}{{\rd t}}
\nc{\pp}{{p_\mu v^\mu}} \nc{\ppo}{{\po_\mu \gamma^\mu}}
\nc{\ovsbf}{{\overline{\sbf}}}
\nc{\cmf}{_{\star}} \nc{\para}{^{\parallel}} \nc{\orto}{^{\perp}}
\nc{\vi}{{v^{(\rm i)}}} \nc{\aii}{{a^{(\rm 2i)}}}
\nc{\M}{{\rm I\!\!M}} \nc{\ain}{{a^{(\rm 2n)}}}
\nc{\impo}{\widehat{\imp}}  \nc{\Ho}{\widehat{H}}
\nc{\doS}{\dot{S}} \nc{\ddoS}{\ddot{S}} \nc{\doJ}{\dot{J}}
\nc{\doL}{\dot{L}} \nc{\ddoL}{\ddot{L}}
\nc{\dof}{\dot{f}} \nc{\ddof}{\ddot{f}} \nc{\dddof}{{\stackrel{\ldots}{f}}}
\nc{\dog}{\dot{g}} \nc{\ddog}{\ddot{g}} \nc{\dddog}{{\stackrel{\ldots}{g}}}
\nc{\doabf}{\mbox{\boldmath ${{\stackrel{.}{a}}}$}}
\nc{\ddoabf}{\mbox{\boldmath ${{\stackrel{..}{a}}}$}}
\nc{\dddoabf}{\mbox{\boldmath $\dddoa$}}
\nc{\ddddoabf}{\mbox{\boldmath $\ddddoa$}}
\nc{\CoMF}{_{\rm {\footnotesize CMF}}}

\begin{document}

\title{NON-NEWTONIAN MECHANICS\footnote{Work partially supported by
I.N.F.N. and M.U.R.S.T.}}

\author{GIOVANNI \ SALESI}

\date{}
\maketitle
\begin{center}
{{\em Universit\`a Statale di Bergamo, Facolt\`a di Ingegneria, Italy};\\
and\\
{\em Istituto Nazionale di Fisica Nucleare--Sezione di Milano, Italy$^{\star}$
}}\\
\end{center}
\footnotetext{$\!\!\!\!^{\star}\,$e-mail: {\em salesi@ct.infn.it}}

\vs{0.5 cm}

\noindent The classical motion of spinning particles can be described without
recourse to particular models or special formalisms, and without employing
Grassmann variables or Clifford algebras, but simply by generalizing
the usual spinless theory. We only assume the invariance with respect
to the Poincar\'e group; and only requiring the conservation of the linear and
angular momenta we derive the {\em zitterbewegung}, namely the decomposition of
the \mbox{4-velocity} in the usual newtonian constant term $p^\mu/m$ and in a
non-newtonian time-oscillating spacelike term. Consequently, free classical
particles do not obey, in general, the Principle of Inertia. Superluminal
motions are also allowed, without violating special relativity, provided that
the energy-momentum moves along the worldline of the center-of-mass. Moreover,
a non-linear, non-constant relation holds between the time durations measured
in different reference frames. Newtonian mechanics is re-obtained as a particular
case of the present theory: namely for spinless systems with no zitterbewegung.
Then we analyze the strict analogy between the classical zitterbewegung
equation and the quantum Gordon-decomposition of the Dirac current. It is
possible a variational formulation of the theory, through a Lagrangian
containing also derivatives of the 4-velocity: we get an equation of the
motion, actually a generalization of the Newton law $a=F/m$, where
non-newtonian zitterbewegung-terms appear. Requiring the rotational symmetry
and the reparametrization invariance we derive the classical spin vector and
the conserved scalar Hamiltonian, respectively. We derive also the classical
Dirac spin $(\abf\times\vbf)/4m$ and analyze the general solution of the
Eulero-Lagrange equation oscillating with the Compton frequency $\omega=2m$.
The interesting case of {\em spinning} systems with zero intrinsic
angular momentum is also studied.


\newpage

{\sl ``If a spinning particle is not quite a point particle, nor
a solid three  dimensional top, what can it be?''}

\rightline{Asim O. Barut \qquad \qquad \qquad \qquad}
\section{Classical non-newtonian systems} 

\subsection{Introduction} 

\noindent The theory we are going to put forward concerns {\em classical
systems} (CS's) in the most general meaning of the word, namely {\em
non-quantum systems}. As special relativity allows, a CS can own ``internal''
degrees of freedom and a spin angular momentum. The set of CS's contains as a
special subset, the one of {\em spinless systems}, that we shall call
also {\em newtonian systems} (NS's), such as, e.g., macroscopic bodies
studied in newtonian classical mechanics.
Since the works by Compton\cite{Compton}, Uhlenbeck and Goudsmith\cite{Uhlenbeck},
and Frenkel\cite{Frenkel}, many classical theories of
spinning particles have been investigated for about eighty years\cite{Classical}.
Grassmann variables in classical actions for spinning systems has been
employed by Berezin and Marinov\cite{Berezin}, Ikemori\cite{Ikemori} and
Casalbuoni\cite{Casalbuoni}.
In the last twenty years a renewed interest has arisen towards classical
theoretical approaches to microsystems, especially in applications to
(super)strings and membranes, in view of a possible unification of the
elementary forces of Nature. In this section we shall obtain important
properties, constraints and equations
which rule the kinematics of free classical particles endowed with spin,
without any recourse to ad-hoc theories or additional assumptions besides the
requirements of the usual spacetime symmetries, and without any recourse to the
non-commuting numbers of the Grassmann algebra or to the multivectors of
the Clifford ``Spacetime'' algebra\cite{Hestenes}.

\

\noindent In the absence of external fields the spacetime isotropy
implies the conservation of the total angular momentum:
\bb
\doJ^\munu = \doL^\munu + \doS^\munu = 0\,,                \label{eq:A}
\ee
tensor $L^\munu\equiv x^\mu p^\nu-x^\mu p^\nu$ being the orbital angular
momentum, and tensor $S^\munu$ the spin angular momentum.
The derivation is taken with respect to the proper time $\tau$, defined as
{\em the time measured in the center-of-mass frame} (CMF) where, by definition,
the \mbox{3-momentum} vanishes, $\imp=0$. The adopted metric is $(+;\;-,-,-)$.
The symmetry under spacetime translations involves the
conservation of the 4-momentum $p^\mu \equiv (p^0;\;\; \imp)$\,:
\bb
\dop^\mu \ug 0\,.                                          \label{eq:B}
\ee
We want to stress that the above conservation laws will be sufficient
to derive all the equations and the constraints of the motion.
The consequent theory will be the most general one and will not be a result of
a particular theoretical model adopted. Hereafter we shall choose units such
that $c=1$. Being \ $L^\munu \equiv x^\mu p^\nu-x^\nu p^\mu$, \
from (\ref{eq:A}) and (\ref{eq:B}) we have:
\bb
\doS^\munu = -\doL^\munu = p^\mu v^\nu-p^\nu v^\mu\,,    \label{eq:C}
\ee
where, as usual, the 4-velocity is defined as the proper-time derivative of the spacetime
coordinate:
\bb
v^\mu \equiv \dox^\mu \equiv \left(\frac{\dt}{\dtau};\;\frac{\rd\xbf}{\dtau}
\right)\,.
                                                           \label{eq:E}
\ee
We are not forced, because of mathematical or physical reasons, to assume
a priori that the CMF, where $\imp=0$,
must coincide with the reference system where \mbox{$\vbf\equiv\rd\xbf/\dtau=0$},
namely the {\em rest frame} (RF), where by definition the speed vanishes.
Then, except particular initial or boundary conditions, in general we can write:
\bb
\vbf\CoMF \equiv \left.\frac{\rd\xbf}{\dtau}\right|\CoMF \neq 0\,.
                                                           \label{eq:nonquiet}
\ee
On the other hand the above statement agrees with the physical structure of
quantum probability currents and quantum velocity operators for spinning systems,
as we shall later see in subsection 1.4. The Lorentz invariant $v^2={v^0}^2 -
\vbf^2$ can be evaluated in the CMF, where $v^0\CoMF=\dtau/\dtau=1$
identically (hereafter whichever quantity referred to the CMF will be labelled
by $\star\;$):
\bb
v^2 \ug 1 - \vbf^2\cmf\,;                                  \label{eq:II}
\ee
from which, taking in account eq.\,(\ref{eq:nonquiet}),
\bb
v^2\neq 1\,.                                               \label{eq:G}
\ee
It follows that $v^\mu$ cannot be put in the usual form
$$
v^\mu \neq \left(\frac{1}{\sqrt{1-w^2}}; \;\;
\frac{\wbf}{\sqrt{1- w^2}}\right)\,.
$$
Furthermore, $v^2$ is not, a priori, required to be a time-constant quantity.

Let us write down the two basic invariant constraints, found also in classical
theory of spinless systems:
\bb
p^2 \ug m^2\,;                                             \label{eq:H1}
\ee
\bb
\pp \ug m\,.                                               \label{eq:H2}
\ee
The first constraint ---which expresses the conservation of $p^\mu$ given by
eq.\,(\ref{eq:B})--- implies the second one. In fact the relativistic invariant
$\pp \equiv p_0v^0-\imp\cdot\vbf$ is nothing but the energy in the CMF
$p^0\cmf$, because in the CMF $v^0\cmf=1$ and $\imp\cmf=0$ by definition:
\ $\pp=p^0\cmf$. \ Because of (\ref{eq:H1}) $\imp\cmf=0$ implies
$p^0\cmf=\pm m$ \,\, ($m\equiv\sqrt{p^2}>0$);
and, if we choose the positive sign for the CMF energy, $\pp$ results to be
equal just to $m$.

Before going on, we want to remark that the wide generality and novelty of the
results we shall obtain is due in particular to our assumption that {\em the
proper time is the time elapsed in the CMF and not in the RF}, and then that
\ $v^2\neq 1$. By contrast, in the literature (with some exception as the
Barut--Zanghi model\cite{Barut})
the proper time $\tau$ is defined as the RF time, $\tau\equiv t_{\rm RF}$. In
the latter case we have as usual $v^2=1$ in any frame, since
\ $v^2\equiv(\dt/\dtau)^2 - (\rd\xbf/\dtau)^2$ \ is actually equal to 1 in the
RF, where by definition $\rd\xbf/\dtau=0$. Notice also that the above
deduction of (\ref{eq:H2}) from (\ref{eq:H1}) does not hold anymore with such a
definition of $\tau$, since now \ $v^0\cmf\equiv\dt\cmf/\dtau\neq 1$, \ and
then
\bb
\pp\equiv mv^0\cmf\equiv\Mc\neq m\,.                       \label{eq:pv}
\ee
Notwithstanding, in literature both constraints $\pp=m$ and $v^2=1$, mutually
excluding for particles endowed with spin, are simultaneously assumed.

\subsection{Zitterbewegung} 

Let us come back to our proper time approach, $\tau\equiv t_{\rm CMF}$. By
multiplying both sides of (\ref{eq:C}) times $p_\nu$ and exploiting conditions
(\ref{eq:H1}) and (\ref{eq:H2}), we derive out\footnote{Another zitterbewegung
equation similar, but {\em not equivalent} to (\ref{eq:I}), is the well-known
{\em Corben--Papapetrou equation}\cite{Corben, Papapetrou} $v^\mu \ug p^\mu/\Mc
- \doS^\munu v_\nu/\Mc$, where $\Mc\equiv \pp\neq m$. We have also to
account that in Corben's theory any derivative is taken with respect to the RF time, and
not to the CMF time, as in the present approach. Notice also that, by
contrast with (\ref{eq:I}), in the Corben--Papapetrou equation the zitterbewegung term is
{\em not} in general orthogonal to the newtonian term $p^\mu/m$.
Obviously, for spinless NS's which do not show zitterbewegung,
$\doL^\munu = \doS^\munu = 0$, both equations of the motion
reduce to the usual newtonian relation $v^\mu = p^\mu/m$, with $v^2=1$.}
\bb
\fbox{${\dis v^\mu \ug \frac{p^\mu}{m} - \frac{\doS^\munu p_\nu}{m^2}}$}\,.
                                                           \label{eq:I}
\ee
The above equation can be re-written also in terms of orbital angular
momentum rather than of spin tensor:
\bb
v^\mu \ug \frac{p^\mu}{m} + \frac{\doL^\munu p_\nu}{m^2}\,.
\ee
The peculiar occurrence that in general the velocity is not constant and not
parallel to the momentum, is the so-called {\em zitterbewegung}\cite{Zbw,
Salesi, Schroedinger}. We then shall call eq.\,(\ref{eq:I}) {\em
zitterbewegung equation for a free particle}.
The global velocity contains a ``translational'', ``newtonian'', time-constant
component $p^\mu/m$ related to the motion of the CM; and a ``rotational'',
``non-newtonian'', time-varying component related to the presence of the spin.
As a consequence the RF, where $\vbf=0$, and the CMF, where $\imp=0$, in general
do not coincide. In particular, the presence of zitterbewegung implies a
motion even in the CMF, and then in the {\em non-relativistic limit}: in fact, for
$\imp\to 0$ we have $\doS^{ik}\to 0$ (since the spin 3-vector conserves in
non-relativistic mechanics) but $\doS^{i0}{\to\!\!\!\!\!\!/} \ \ 0$ ($S^{i0}$
is not required to conserve), so that from eq.\,(\ref{eq:I}) we have
$v^i\to-\doS\cmf^{i0}/m\neq 0$.

\

\noindent From the equation (\ref{eq:I}) it follows that in a generic frame
the trajectory will be a
helix around the constant direction of $\imp$. Notice that, because identically
\ $\doS^\munu p_\nu p_\mu=0$ \ (the contraction of an antisymmetric tensor with a
symmetric tensor always vanishes), {\em the spin term in the global velocity
results always spacelike and orthogonal to the timelike Newtonian
component} $p^\mu/m$. This property recalls
the known dispersion relation $|\vbf||\Vbf|=c^2$ between the timelike
(external) group-speed and the spacelike (internal) phase-speed found, for
instance, in de Broglie's ``pilot-wave'' (``double-solution'') theory.
From (\ref{eq:I}) we see that the {\em zitterbewegung originates from the
non-conservation of the orbital angular momentum and of the spin angular
momentum} (even if their sum conserves)
\bb
\doS\neq 0 \qquad\qquad\qquad \doL\neq 0\,.
\ee
Let us underline that in some papers
$S^\munu p_\nu=0$ is arbitrarily assumed, so that \mbox{$\doS^\munu
p_\nu=0$:}
i.e., for (\ref{eq:I}), no zitterbewegung, in spite of the presence of
spin. Here we will not in any way limit the generality of the theory and shall
not make further assumptions: And in section 4 we shall see that the
motion of classical Dirac particles undergoes zitterbewegung with
$S^\munu p_\nu=a^\mu/4\neq0$, see eq.\,(\ref{eq:doS}).
Derivating both sides of eq.\,(\ref{eq:I}) we get
\bb
a^\mu \ug -\,\frac{\ddoS^\munu p_\nu}{m^2}\,,              \label{eq:QQ}
\ee
or also
\bb
a^\mu \ug \frac{\ddoL^\munu p_\nu}{m^2}\,.
\ee
Therefore, while for NS's $a^\mu=0$ in the absence of external forces,
for CS's in general $a^\mu\neq 0$ so that {\em the Galileo-Newton Principle
of Inertia does not hold anymore}.\\
\noindent From (\ref{eq:C}) we obtain:
$$
\doS^\oi \ug p^0v^i-p^iv^0\,,\qquad\qquad (i=1,2,3)
$$
and then
$$
v^i \ug \frac{p^iv^0}{p^0} +
\frac{\doS^\oi}{p^0}\,.                            \label{eq:F}
$$
Dividing both sides for $v^0$, and taking in account
(\ref{eq:E}) and the identity
$$
\frac{v^i}{v^0}\equiv
\left(\frac{\rd x^i}{\dtau}\right)\left(\frac{\dt}{\dtau}\right)^{-1}
\equiv\frac{\rd x^i}{\dt}\,,
$$
we get
\bb
\frac{\rd x^i}{\dt} \ug \frac{p^i}{p^0} + u^i\,,   \label{eq:V}
\ee
where $u^i\equiv\doS^\oi/v^0p^0$. \ Notice that, whilst the speed $|\imp|/p^0$
of the CM is always smaller than the speed of light in vacuum $c$, the
zitterbewegung speed $|\ubf|$ is not constrained at all
(see below). Therefore, {\em without violating special relativity, we can
allow superluminal motions} of spinning charges, provided that signals and
momenta are carried by the CM (it follows also that the reference systems, as
expected, are endowed with subluminal relative speeds).

\subsection{General properties of the inertial motion} 

\subsubsection{Constraints on $v^2$ and motions in the CMF} 

Let us write the zitterbewegung equation eq.\,(\ref{eq:I}) in a compact form
\bb
v^\mu \ug w^\mu + V^\mu\,,                                 \label{eq:PP}
\ee
where $w^\mu\equiv(1/\sqrt{1-w^2};\;\wbf/\sqrt{1-w^2})\equiv p^\mu/m$
and $V^\mu\equiv-\doS^\munu p_\nu/m^2$\footnote{Even if endowed with different
transformation properties, the space parts of 4-vectors (as, e.g., $\vbf$,
$\imp$, $\xbf$) and the 3-vectors (as, e.g., $\wbf$ or the spin vector $\sbf$)
will be for convenience labelled by means of boldface symbols.}. \ Since, as already seen,
\bb
w_\mu V^\mu=-\doS^\munu p_\nu p_\mu/m^3=0                 \label{eq:orto}
\ee
(4-orthogonality between newtonian and non-newtonian terms in $v^\mu$) and
$w^2=p^2/m^2=1>0$, we have:
\bb
V^2\leq 0\,.                                               \label{eq:V2first}
\ee
[On the other hand we have identically $v^0\cmf=w^0\cmf=1$ which for
(\ref{eq:PP}) implies
\bb
V^0\cmf=0\,,                                               \label{eq:V0}
\ee
which in its turn involves just $V^2=0-\Vbf^2\cmf\leq 0$]. Because of
(\ref{eq:orto}), (\ref{eq:V2first}), and of the decomposition $v^2 = w^2 + V^2
+ 2\,w_\mu V^\mu = 1 + V^2$, the following constraint holds:
\bb
-\infty < v^2 \leq 1\,.                                    \label{eq:v2<1}
\ee
If, at a given time, $0<v^2=1-\vbf\cmf^2<1$ ({\em timelike} case),
the corresponding motion is subluminal in the CMF, in that $\vbf\cmf^2<1$. \
If instead we have $v^2<0$ ({\em spacelike} case) the motion is superluminal,
``tachyonic'', since $\vbf\cmf^2>1$. \ In the special {\em lightlike} case,
$v^2=0$, we have $\vbf\cmf^2=1$, and the charge moves in the CMF at the speed
of light $c$.\footnote{Let us suppose that for all the massive elementary (not
composed) particles, i.e. for electrons and quarks, it {\em always} is $v^2=0$.
We might therefore state that $c$ must not be meant as the {\em maximum} speed,
but actually as the {\em unique} speed of any pointlike charge.
In such a way eachever subluminal speed observed ---energy and momentum travel
at a slower-than-light speed $w$--- is to be realized as the component parallel
to the momentum of the total velocity (obviously smaller than the modulus $c$ of
the total velocity).}

By a little algebra we can also obtain the following relations:
\bb
v^2 \ug 1 + \frac{\doS^\munu\doS_\munu}{2m^2}\,,
\ee
\bb
v^2 \ug 1 + \frac{\doS^\munu p_\mu v_\nu}{m^2}\,.         \label{eq:V2last}
\ee

\subsubsection{``Longitudinal'' and ``intrinsic''
zitterbewegung} 

Eq.\,(\ref{eq:orto}) implies $\imp\cdot\Vbf-p_0V^0=0$; then we have orthogonality
between $\imp$ and $\Vbf$, $\imp\cdot\Vbf=0$, only in all those frames where
$$
V^0=0\,.
$$
The reference frames where the time-component of a given spacelike 4-vector
$A^\mu$ vanishes are named ``standard frames for $A^\mu$''\cite{Digiacomo}.
Therefore we have $\imp\cdot\Vbf=0$ only in the standard frames for $V^\mu$,
among which the CMF is a particular case (since $V^0\cmf=0$,\, eq.\,(\ref{eq:V0})).
In these reference systems it holds a particular case of the general constraint
$p_\mu a^\mu=0$ [obtained by time-derivating side by side eq.\,(\ref{eq:H2})],
namely: $\dot{V}^0=a^0=\imp\cdot\abf=0$. The considered frames can be obtained
by applying to the CMF a Lorentz boost $\wbf$ orthogonal to the zitterbewegung
plane, $\wbf\cdot\Vbf\cmf=0$. In such a way, as it is easy to see, the
spacelike 4-vector $V^\mu$, equal in the CMF to $(0; \Vbf\cmf)$, will transform
in itself (namely, it is an eigenvector of the matrix operating the considered
Lorentz transformation), so that $V^0$ still vanishes.\footnote{By contrast,
anyever Lorentz boost changes the time-component of a generic {\em timelike}
4-vector.} In section 4, in studying the classical Dirac theory, we shall see
that the standard frames, in which we observe a pure transverse
($\imp\cdot\Vbf=0$) zitterbewegung, are the frames in which it appears as a
polarized particle, i.e., with the spin aligned along the momentum. By
contrast, {\em in a generic frame we have also a component of the zitterbewegung
parallel to the momentum, with} (cf. section 4) {\em oscillations of the charge
along the rectilinear trajectory of the} CM (``longitudinal'' or ``extrinsic''
zitterbewegung), besides the oscillations transverse to the momentum
(``transverse'' or ``intrinsic'' zitterbewegung).

\subsubsection{Non-constant times-ratio} 

Quantity $v^0=w^0+V^0=\dt/\dtau$ may be defined as ``times-ratio'', in that
measures the ratio between the time durations referred to a generic reference
system ($\dt$) and to the CMF ($\dtau$). It generalizes the concept of
Lorentz factor in the present theory. But, whilst for NS's
$v^0=w^0=\gamma\equiv 1/\sqrt{1-w^2}$ is
always a constant quantity due to (\ref{eq:B}), for generic CS's $v^0
\neq 1/\sqrt{1-w^2}$ is time-varying (and in particular time-oscillating,
cf.\,section 4) since $V^0$ is not forced to be a constant quantity. Thus,
{\em the times-ratio is not time-constant} anymore, as it instead occurs in
special relativity for spinless NS's. In a sense, we might speak of a {\em
non-constant Lorentz factor}. Moreover, the times-ratio is not necessarily
larger than 1: we may also have a time-contraction, besides the usual
time-dilation, see section 4. (By contrast, in the standard frames for $V^0$
we have $V^0=0$, and then the times-ratio turns out to be the usual constant
quantity $\gamma$\footnote{This result can be alternatively derived
by considering equation (\ref{eq:H2}) $\pp=m$. In fact, in the standard frames
we have $\imp\cdot\Vbf=0$, which, for (\ref{eq:PP}), implies
$\imp\cdot\vbf=\imp\cdot\wbf/\sqrt{1-w^2}$. It follows that
$\pp\equiv p_0v^0-\imp\cdot\vbf=p_0v^0-\imp\cdot\wbf/\sqrt{1-w^2}=m$, from
which (exploiting also $p^\mu\equiv mw^\mu$) we get the constant times-ratio
$v^0=1/\sqrt{1-w^2}$.}). In general it is easy to see that {\em a non-linear
relation occurs between the time durations measured in different reference
systems} (see section 4).

Being all four components of $v^\mu$ not constant, we may also say that
the trajectory is a helix not only in the ordinary space $\R^3$: also in the
Minkowski spacetime $\M^4$ the trajectory is a 4-dimensional helix spyralizing
around the constant 4-vector $p^\mu$.

\subsection{Correspondences between the classical velocity and the quantum
probability current} 

The most impressive correspondence between the present classical theory and
the standard wave-mechanics may be found in the celebrated {\em Gordon
decomposition} of the conserved current of the Dirac equation\cite{Gordon},
which writes (hereafter we assume $\hbar=1$):
\bb
j^\mu = \psib\gm\psi = \frac{1}{2m}\,[\psib(\po^\mu\psi) - (\po^\mu\psib)\psi]
+ \frac{1}{m}\pa_\nu\,(\psib\So^{\mu\nu}\psi)\,,          \label{eq:S}
\ee
where $\psib\equiv\psid\gao$ is the usual hermitian-adjoint bispinor,
$\po^\mu \equiv i\pa^\mu$ is the 4-dimensional momentum operator, and
$\So^{\mu\nu}\equiv i(\gm\gn - \gn\gm)/4$ represents the spin-tensor operator.
In fact, the standard interpretation of the above decomposition
quite agrees with our zitterbewegung equation
(\ref{eq:I}). The first term in the r.h.s. is associated with the
translational motion of the CM (the scalar part of the current,
corresponding to the Klein--Gordon current). As a matter of fact,
for a momentum eigenstate, i.e. for a plane-wave, this term turns out to be
proportional to $p^\mu/m$. \ By contrast, the non-newtonian term in the r.h.s.
is related to the existence of the spin, and describes the zitterbewegung
rotational motion. The correspondences and analogies between classical and
quantum laws do not concern only the probability current, but concern also the
operators of the basic kinematic quantities. In Dirac theory, indeed, both
the 4-velocity operator $\gm$ and the 3-velocity operator
$\albf\equiv\gao\gabf$ do not commute with the Dirac Hamiltonian
$\Ho=\albf\cdot\impo+m\gao$. Therefore, such as it happens for CS's, also in
quantum mechanics those quantities, differently from the momentum, are
{\em not} time-constant. Let us recall that the zitterbewegung actually occurs also
for non-relativistic particles, in the framework of the Pauli and Schr\"odinger
theories\cite{Hestenes, Salesi, Holland}.
In fact, following Landau\cite{Landau-1}, we can write a non-relativistic
Gordon-like decomposition of the conserved Pauli current
\bb
\jbf = \frac{i}{2m}[(\vecna\psid)\psi - \psid\vecna\psi]
+ \frac{1}{m}\vecna\times(\psid\sigbf\psi)\,,
\ee
where $\psi$ is a Pauli 2-components spinor and $\sigbf$ is the usual Pauli
vector (2$\times$2) matrix. Also the above current appears as a sum of a
newtonian part which, at the classical limit ($\hbar\to 0$), is
parallel to the classical momentum (equal to $\hbar$ times the gradient of the
action); and of a non-newtonian part due to the spin, which vanishes only
at the classical limit, i.e. for spinless bodies, {\em but not in the
non-relativistic limit}, i.e. for a small momentum $\imp$.

Analogous Gordon-like decompositions of the conserved 4-currents can be found
also for spin-1 bosons and for spin-${3\over 2}$ fermions in the Proca and
Rarita-Schwinger theories, respectively.

\section{Lagrangian theory} 

\subsection{Generalized Newton equation} 

\noindent A Lagrangian for a free spinless NS, which must be invariant under
the Lorentz group as well as under the space and time inversions,
can depend only on the 4-velocity squared and is often written in the
following form\cite{Landau-2} (hereafter a generic 4-vector $a^\mu$ is for
simplicity indicated only by $a$, and a scalar product $a_\mu b^\mu$ by $ab$)
\bb
\Lc \ug {1\over 2}\,m\,v^2\,.
\ee
Searching the minimum of the action ${\cal S}=\int\!\!\Lc\,\dtau$ leads to the
Eulero-Lagrange equation
$$
\frac{\pa\Lc}{\pa x} = \dot{\frac{\pa\Lc}{\pa\dox}}\,,
$$
which implies the usual rectilinear uniform motion of free newtonian bodies
\bb
a = 0\,.
\ee
The corresponding momentum is
\bb
p \equiv \frac{\pa\Lc}{\pa\dox} \ug m\,v\,,
\ee
with ---as expected for NS's--- no zitterbewegung. By contrast, the general
inertial motion of a CS is endowed with zitterbewegung and non-constant
velocity. Actually, a priori, the derivatives of the velocity do not vanish.
By requiring the symmetry under the Poincar\'e group, the Lagrangian can
depend, besides on the 4-velocity squared, also on its (any order) derivatives
squared. Therefore a Lagrangian for a CS may be taken as follows:
\bb
\Lc \equiv \um\,m\,v^2 + \um k_1\,\dov^2 + \um k_2\,\ddov^2 + \cdots
\equiv \sum_{i=0}^\infty\,\um k_i\,\vi^2\,,
\ee
where the $k_i$ are constant scalar coefficients, $k_0=m$, and
$$
\vi \equiv \frac{\rd^{\rm i}v}{\dtau^{\rm i}}\,.
$$
As are going to see, the above Lagrangian will imply the expected 4-velocity,
sum of a translational part and of a spin part. In the presence of an external
force the generalization can be made as usual, with the introduction of a
scalar potential $U(x)$
\bb
\Lc \equiv \sum_{i=0}^\infty\,\um k_i\,\vi^2 - U\,.        \label{eq:Lc}
\ee
The coefficients $k_i$ ---which may be chosen equal to zero for $i$ larger than
a given integer, see below--- might be functions of the self-interaction of
the particle and of its mass and charge: Let us recall, for comparison, the
well-known infinite-terms equation of the self-radiating classical electron or
the ``cronon'' theory of the electron (reviewed at the end of the present
section). In other theoretical frameworks, the coefficients $k_i$ can be
related to the underlying string structure (or membrane or $n$-brane structure)
of a spinning particle. Polyakov and others\cite{Polyakov, Pavsic} have
proposed a classical string action in which, besides the ordinary Nambu--Goto
term, appear additional terms dependent on the so-called ``rigidity'' or on the
so-called ``extrinsic curvature'': then on the 4-acceleration squared. Classical
equations of the motion for a rigid $n-$dimensional worldsheet, either in flat
or curved background spacetimes, have been derived from Lagrangians containing
also terms dependent on higher derivatives of the 4-velocity (``torsion''-terms,
etc.).

The Eulero-Lagrange equation for a generic Lagrangian $\Lc(x,\dox,\ddox,
\ldots)$ derived from the Principle of Least Action is
\bb
\frac{\pa\Lc}{\pa x} = \dot{\frac{\pa\Lc}{\pa\dox}} -
\ddot{\frac{\pa\Lc}{\pa\ddox}} + {\stackrel{\ldots}{\frac{\pa\Lc}{\pa\dddox}}} -
\cdots\,.                                                  \label{eq:EL}
\ee
From eqs.(\ref{eq:Lc}) and (\ref{eq:EL}) we can write the {\em generalized
Newton equation of the motion}:
\bb
\fbox{${\dis-\,\frac{\pa U}{\pa x} \ug m\,a - k_1\,\ddoa + k_2\,\ddddoa - \cdots\equiv
\sum_{i=0}^\infty\,(-1)^{{\rm i}}k_i\,\aii}$}\,.
                                                           \label{eq:GNEq}
\ee
The 4-momentum $p$ is that quantity which is conserved under 4-translations
for free systems ($\dop=0$ if $U=0$) and whose time derivative is the
4-force $F$:
\bb
\dop = \frac{\pa\Lc}{\pa x} = -\,\frac{\pa U}{\pa x} = F\,.  \label{eq:dop}
\ee
The requirement of spacetime homogeneity
$$
\frac{\pa\Lc}{\pa x} \ug 0
$$
because of eq.\,(\ref{eq:EL}) implies
$$
0 = \dot{\frac{\pa\Lc}{\pa\dox}} - \ddot{\frac{\pa\Lc}{\pa\ddox}} +
{\stackrel{\ldots}{\frac{\pa\Lc}{\pa\dddox}}} - \cdots =
\frac{\rd}{\rd\tau}\left[\frac{\pa\Lc}{\pa\dox} - \dot{\frac{\pa\Lc}{\pa\ddox}} +
\ddot{\frac{\pa\Lc}{\pa\dddox}} - \cdots\right]\,.
$$
Then the quantity in square brackets is the conserved momentum of a CS:
\bb
p \ug \frac{\pa\Lc}{\pa\dox} - \dot{\frac{\pa\Lc}{\pa\ddox}} +
\ddot{\frac{\pa\Lc}{\pa\dddox}} - \cdots                   \label{eq:Mom}
\ee
which, for Lagrangians (\ref{eq:Lc}) can be written as
\bb
p = m\,v - k_1\,\ddov + k_2\,\ddddov - \cdots
\equiv\sum_{i=0}^\infty\,(-1)^{{\rm i}}\,k_i\,v^{({\rm 2i})}\,.
\ee
The zitterbewegung part of the velocity $v_{\rm zbw}=v-p/m$ reads:
\bb
v_{\rm zbw} = \frac{1}{m}\,\left(\,k_1\,\ddov - k_2\,\ddddov + \cdots\right)
\equiv -\,\frac{1}{m}\,\sum_{i=1}^\infty\,(-1)^{{\rm i}}\,k_i\,v^{({\rm 2i})}\,.
\ee
The orbital angular momentum of a CS is the sum of the usual newtonian term and
of a non-newtonian zitterbewegung term
\bb
\Lbf = \xbf\times\imp = \xbf\times m\vbf +
\sum_{i=1}^\infty\,(-1)^{{\rm i}}\,\xbf\times k_i\,\vbf^{({\rm 2i})}\,.
\ee
For free particles ($U=0$) the generalized Newton equation (\ref{eq:GNEq})
reduces to:
\bb
0 \ug m\,a - k_1\,\ddoa + k_2\,\ddddoa - \cdots\,.      \label{eq:NT}
\ee

\

\

\noindent As far as we know, only Caldirola's classical theory of the electron
---based on the existence of an elementary time duration: the
``cronon''\cite{Caldirola}---
is an infinite-order Lagrangian of the same kind of (\ref{eq:Lc}).
In the cronon theory, which applies to charged leptons, the constant
coefficients of the derivatives of the velocity are linked to the electrical
charge $e$:
\bb
k_i \equiv (-1)^i\,\frac{m\,T^{2i}}{(2i+1)!}\,,            \label{eq:ki}
\ee
where $T$ is the already mentioned cronon
$$
T\equiv \frac{4}{3}\,\frac{e^2}{mc^3}\,.
$$
According to this choice, and assuming $U\equiv eA_\mu v^\mu$, the Eulero-Lagrange
equation results to be the following {\em finite-differences equation}
(herafter we come back to the previous asssumption $c=1$)
$$
m\,\frac{v^\mu(\tau+T) - v^\mu(\tau-T) + v^\mu(\tau)\,v^\nu(\tau)\,[v_\nu(\tau+T)
-v_\nu(\tau-T)]}{2T} = e\,F^\munu(\tau)v_\nu(\tau)\,,
$$
which appears as a (non-newtonian) time-symmetrical discretization of the
Lorentz non-radiating equation for the motion of a (newtonian) spinless charge
$$
m\,\dov^\mu\ug e\,F^\munu v_\nu\,.
$$
The cronon theory is rather interesting, among other things, for it seems to
overcome well-known problems due to the electric self-interaction as the
so-called ``runaway solutions'' of the Lorentz-Dirac equation of the electron.
Moreover, Caldirola's theory seems to explain the origin of the
``classical (Schwinger's) part'', $e\hbar/2mc\cdot\alpha/2\pi=e^3/4\pi mc^2$
of the anomalous magnetic momentum of the electron as well as the mass spectrum
of charged leptons. Because of the classical, non-quantum character of his
theory, Caldirola excluded a priori the existence of
spin contributions or zitterbewegung terms in its theory. We instead are going
to show, in the next subsection, the arising of a very intrinsic angular
momentum for any given $\Lc$.

\

\noindent For each finite $n$, we define a $n$-th order Lagrangian as follows:
\bb
\Lcn \equiv \sum_{i=0}^n\,\um k_i\,\vi^2\,,                \label{eq:Lcn}
\ee
and look for the consequent motions of the system. The generalized Newton
equation (\ref{eq:NT}) is now a linear constant-coefficients differential
equation of $n$-th order:
$$
0 \ug m\,a - k_1\,\ddoa + k_2\,\ddddoa - \cdots +
(-1)^{{\rm n}}k_n\,a^{{\rm (2n)}}\,.
$$
The associated ``characteristic equation'' is
$$
0 \ug m - k_1\,z^2 + k_2\,z^4 - \cdots + (-1)^{{\rm n}}\,k_n\,z^{2n}\,.
$$
If we forbid exponentially spreading or collapsing motions, but, on the
contrary, we ask finite,
periodic zitterbewegung motions (around the uniform translation of the CM)
the signs of the coefficients $k_i$ must be alternate. In fact, to have only
oscillating motions, each solution ${z_i}^2$ ($i=1,\,\ldots\,n$) of the
characteristic equation must satisfy
$$
z_i^2 < 0\,,
$$
since $z_i^2=-\omega_i^2$, where $\omega_i$ is the $i$-th frequency
of the motion.
Therefore, because of the Descartes rule, we have to ask
\bb
{\rm sign}(k_i) \ug (-1)^{{\rm i}}\,.
\ee
Notice that the above condition is satisfied in the cronon theory,
eq.\,(\ref{eq:ki}), as well as in our future applications: see, e.g.,
sections 3 and 4.

\subsection{Classical spin} 

\noindent Any Poincar\'e-invariant Lagrangian $\Lc$\footnote{Let us underline
that each result obtained in this subsection holds not only for free systems,
but equally in the presence of an external {\em scalar} potential $U$.}
is also invariant under the 4-rotations group. Then, for
the N\"other theorem, the angular momentum tensor $J^\munu$ is conserved.
For the deduction of $J^\munu$, we firstly work supposing only that
our $\Lc$ be a Poincar\'e-invariant function of $x$ and of its time derivatives
\bb
\Lc \equiv f(\tau; x,v,\dov,\ddov,\ldots)\,,              \label{eq:Lf}
\ee
and without recourse to the more specific form given by (\ref{eq:Lc}). We then
ask that the Lagrangian after an infinitesimal 4-rotation of the reference frame
does not vary:
\bb
0 = \delta\Lc =
\frac{\pa\Lc}{\pa x}\delta x +
\frac{\pa\Lc}{\pa v}\delta v +
\frac{\pa\Lc}{\pa \dov}\delta \dov +
\frac{\pa\Lc}{\pa \ddov}\delta \ddov +
\frac{\pa\Lc}{\pa \dddov}\delta \dddov + \cdots\,.
\ee
If $\delta\Omn$ is the antisymmetric 4-tensor giving the infinitesimal
rotation angles around an axis orthogonal to the plane $x^\mu x^\nu$
we shall have
\bb
0 = \delta\Lc =
\frac{\pa\Lc}{\pa x^\mu}\delta\Omn\!x_\nu +
\frac{\pa\Lc}{\pa v^\mu}\delta\Omn\!v_\nu +
\frac{\pa\Lc}{\pa \dov^\mu}\delta\Omn\!\dov_\nu +
\frac{\pa\Lc}{\pa \ddov^\mu}\delta\Omn\!\ddov_\nu +
\frac{\pa\Lc}{\pa \dddov^\mu}\delta\Omn\!\dddov_\nu + \cdots\,.
                                                           \label{eq:delta}
\ee
If we consider the first-order Lagrangian \ $\Lc(\tau; x,v)$, i.e.,
the usual newtonian Lagrangian, we can write, being
\ $p_\mu=\pa\Lc/\pa v^\mu$ \ and \ $\pa\Lc/\pa x^\mu=\dop_\mu$, \
$$
0 = \delta\Lc = \frac{\pa\Lc}{\pa x^\mu}\delta\Omn\!x_\nu +
\frac{\pa\Lc}{\pa v^\mu}\delta\Omn\!v_\nu  =
\dop_\mu\delta\Omn\!x_\nu + p_\mu\delta\Omn\!v_\nu  =
$$
$$
= \um\delta\Omn\![(\dop_\mu x_\nu - \dop_\nu x_\mu) + (p_\mu v_\nu - p_\nu v_\mu)]
= \um\delta\Omn\!\frac{\rd}{\dtau}\left(p_\mu x_\nu - p_\nu x_\mu\right)
$$
where we have exploited the antisymmetry of $\Omn$ and the derivation rule
\bb
\frac{\rd \ }{\dtau}\,(fg) \ug \dof g+f\dog\,.                 \label{eq:DR}
\ee
For the arbitrariness of $\delta\Omn$ we eventually obtain the conservation
of the total angular momentum
$$
\frac{\rd J}{\dtau}^\munu=0\,,
$$
with
\bb
J^\munu = x^\mu p^\nu - x^\nu p^\mu\,.
\ee
We can work analogously for Lagrangians of the second order \ $\Lc(\tau;
x,v,\dov)$:
$$
0 = \delta\Lc =
\frac{\pa\Lc}{\pa x^\mu}\delta\Omn\!x_\nu +
\frac{\pa\Lc}{\pa v^\mu}\delta\Omn\!v_\nu +
\frac{\pa\Lc}{\pa \dov^\mu}\delta\Omn\!\dov_\nu\,.
$$
Being, eq.\,(\ref{eq:Mom}), \ $p_\mu=\pa\Lc/\pa v^\mu - \rd(\pa\Lc/\pa
a^\mu)/\dtau$ \ and \ $\pa\Lc/\pa x^\mu=\dop_\mu$ \ we can also write
$$
0 = \dop_\mu\delta\Omn\!x_\nu + \left(p_\mu+\dot{\frac{\pa\Lc}{\pa a^\mu}}\right)
\delta\Omn\!v_\nu + \frac{\pa\Lc}{\pa a^\mu}\delta\Omn\!a_\nu =
$$
$$
= \um\delta\Omn\!\left[
(\dop_\mu x_\nu - \dop_\nu x_\mu) + (p_\mu v_\nu - p_\nu v_\mu) +
\left(\dot{\frac{\pa\Lc}{\pa a^\mu}}v_\nu - \dot{\frac{\pa\Lc}{\pa a^\nu}}v_\mu\right) +
\left(\frac{\pa\Lc}{\pa a^\mu}a_\nu - \frac{\pa\Lc}{\pa a^\nu}a_\mu\right)\right] =
$$
$$
= \um\Omn\!\frac{\rd}{\dtau}\left[(p_\mu x_\nu - p_\nu x_\mu) +
\left(\frac{\pa\Lc}{\pa a^\mu}v_\nu - \frac{\pa\Lc}{\pa a^\nu}v_\mu\right)\right]\,.
$$
We again obtain the conservation of the total angular momentum
$$
\frac{\rd J}{\dtau}^\munu=0\,,
$$
with
\bb
J^\munu = [x^\mu p^\nu - x^\nu p^\mu] +
\left[v^\mu\frac{\pa\Lc}{\pa a_\nu} - v^\nu\frac{\pa\Lc}{\pa a_\mu}\right]\,.
                                                           \label{eq:J1}
\ee
Let us consider also the Lagrangians of third and fourth order: \ $\Lc(\tau;
x,v,\dov,\ddov)$, \ $\Lc(\tau; x,v,\dov,\ddov,\dddov)$. \
By some algebra, exploiting eq.\,(\ref{eq:dop}) and the expression
of the canonical momentum (\ref{eq:Mom}), as well as the identities
\bb
\frac{\rd \ }{\dtau}(\dof g-f\dog)=\ddof g-f\ddog \qquad\quad
\frac{\rd \ }{\dtau}(\ddof g-\dof\dog+f\ddog)=\dddof g+f\dddog\,,
                                                           \label{eq:DI}
\ee
from (\ref{eq:delta}) we get the following conserved total angular momentum:
\bb
J^\munu = [x^\mu p^\nu - x^\nu p^\mu] +
\left[v^\mu\frac{\pa\Lc}{\pa a_\nu} - v^\nu\frac{\pa\Lc}{\pa a_\mu}\right] +
\left[\left(a^\mu\frac{\pa\Lc}{\pa\doa_\nu} - a^\nu\frac{\pa\Lc}{\pa\doa_\mu}\right) -
\left(v^\mu\frac{\dot{\pa\Lc}}{\pa\doa_\nu} -
v^\nu\frac{\dot{\pa\Lc}}{\pa\doa_\mu}\right)\right]
\ee
for $\Lc(\tau; x,v,\dov,\ddov)$; \ and
$$
J^\munu = [x^\mu p^\nu - x^\nu p^\mu] +
\left[v^\mu\frac{\pa\Lc}{\pa a_\nu} - v^\nu\frac{\pa\Lc}{\pa a_\mu}\right] +
\left[\left(a^\mu\frac{\pa\Lc}{\pa\doa_\nu} - a^\nu\frac{\pa\Lc}{\pa\doa_\mu}\right) -
\left(v^\mu\frac{\dot{\pa\Lc}}{\pa\doa_\nu} -
v^\nu\frac{\dot{\pa\Lc}}{\pa\doa_\mu}\right)\right] +
$$
\bb
+ \left[\left(v^\mu\frac{\ddot{\pa\Lc}}{\pa\ddoa_\nu} -
v^\nu\frac{\dot{\pa\Lc}}{\pa\doa_\mu}\right) -
\left(a^\mu\frac{\dot{\pa\Lc}}{\pa\ddoa_\nu} -
a^\nu\frac{\dot{\pa\Lc}}{\pa\ddoa_\mu}\right) +
\left(\doa^\mu\frac{\pa\Lc}{\pa\ddoa_\nu} -
\doa^\nu\frac{\pa\Lc}{\pa\ddoa_\mu}\right)\right]
\ee
for $\Lc(\tau; x,v,\dov,\ddov,\dddov)$.\,
Thus for $\Lc(\tau; x,v)$ we have no spin, $J_\munu=L_\munu$, $S_\munu=0$, as
expected for NS's; the first spin term appears for $\Lc(\tau; x,v,\dov)$ where we
have for (\ref{eq:J1})
\bb
\fbox{${\dis S^\munu=v^\mu\frac{\pa\Lc}{\pa a_\nu} -
v^\nu\frac{\pa\Lc}{\pa a_\mu}}$}\,.
                                                           \label{eq:Smunu}
\ee
A more specific form of the spin vector
$\sbf\equiv(S^{23},S^{31},S^{12})$ can be found for the Lagrangians $\Lcn$
given by (\ref{eq:Lcn}). We get:
\bb
\fbox{${\dis\sbf=k_1\,(\vbf\times\abf)}$}                  \label{eq:spin1}
\ee
for $\Lcuno$;
\bb
\fbox{${\dis\sbf=k_1\,(\vbf\times\abf) + k_2\,(\abf\times\doabf - \vbf\times\ddoabf)}$}
                                                           \label{eq:spin2}
\ee
for $\Lcdue$; and
\bb
\fbox{${\dis\sbf=k_1\,(\vbf\times\abf) + k_2\,(\abf\times\doabf - \vbf\times\ddoabf) +
k_3\,(\doabf\times\ddoabf - \abf\times\dddoabf + \vbf\times\ddddoabf)}$}
                                                           \label{eq:spin3}
\ee
for $\Lctre$. And so on for larger $n$.

\subsection{Hamiltonian} 

\noindent Lagrangian (\ref{eq:Lf}) describes free systems: it cannot explicitly depend on
the proper-time $\tau$ (reparametrization invariance); actually, the
Lagrangians $\Lcn$ are not explicit functions of the time parameter $\tau$.
Also in the presence of external forces we generally suppose that the potential
$U$ does not depend on the proper time. Therefore, because of the N\"other
theorem, we can always get out a conserved scalar Hamiltonian. Let us write the
total time derivative  [$(\pa\Lc/\pa a)\,b\equiv(\pa\Lc/\pa a^\mu)\,b^\mu$ as
before]
\bb
\frac{\rd\Lc}{\rd\tau} = \frac{\pa\Lc}{\pa\tau} +
\frac{\pa\Lc}{\pa x}\dox +
\frac{\pa\Lc}{\pa v}\dov +
\frac{\pa\Lc}{\pa a}\doa +
\frac{\pa\Lc}{\pa \doa}\ddoa + \cdots\,.
\ee
Imposing the reparametrization invariance $\pa\Lc/\pa\tau\ug 0$, we get:
\bb
\frac{\rd\Lc}{\rd\tau} =
\frac{\pa\Lc}{\pa x}\dox +
\frac{\pa\Lc}{\pa v}\dov +
\frac{\pa\Lc}{\pa a}\doa +
\frac{\pa\Lc}{\pa \doa}\ddoa + \cdots\,.
\ee
Let us now consider, for brevity, only the Lagrangians of the first 3 orders.
By using the Eulero-Lagrange equation (\ref{eq:EL}), as well as identities
(\ref{eq:DR}) and (\ref{eq:DI}), we can re-write the above equation in the
following forms:
\bb
\frac{\rd\Lc}{\rd\tau} =
\frac{\rd}{\rd\tau}\left(\frac{\pa\Lc}{\pa v}v\right)
\ee
for $\Lc(\tau; x,v)$;
\bb
\frac{\rd\Lc}{\rd\tau} =
\frac{\rd}{\rd\tau}\left(\frac{\pa\Lc}{\pa v}v\right) +
\frac{\rd}{\rd\tau}\left(\frac{\pa\Lc}{\pa a}a -
\frac{\dot{\pa\Lc}}{\pa a}v\right)
\ee
for $\Lc(\tau; x,v,\dov)$;
\bb
\frac{\rd\Lc}{\rd\tau} =
\frac{\rd}{\rd\tau}\left(\frac{\pa\Lc}{\pa v}v\right) +
\frac{\rd}{\rd\tau}\left(\frac{\pa\Lc}{\pa a}a -
\frac{\dot{\pa\Lc}}{\pa a}v\right) +
\frac{\rd}{\rd\tau}\left(\frac{\ddot{\pa\Lc}}{\pa\doa}v +
\frac{\pa\Lc}{\pa\doa}\doa -
\frac{\dot{\pa\Lc}}{\pa\doa}a\right)
\ee
for $\Lc(\tau; x,v,\dov,\ddov)$\,.
Hereby we have the following conserved Hamiltonians
\bb
\Hc \ug \frac{\pa\Lc}{\pa v}v - \Lc                        \label{eq:Ham0}
\ee
for $\Lc(\tau; x,v)$;
\bb
\Hc \ug \frac{\pa\Lc}{\pa v}v +
\left(\frac{\pa\Lc}{\pa a}a - \frac{\dot{\pa\Lc}}{\pa a}v\right)
- \Lc                                                      \label{eq:Ham01}
\ee
for $\Lc(\tau; x,v,\dov)$;
\bb
\Hc \ug \frac{\pa\Lc}{\pa v}v +
\left(\frac{\pa\Lc}{\pa a}a - \frac{\dot{\pa\Lc}}{\pa a}v\right) +
\left(\frac{\ddot{\pa\Lc}}{\pa\doa}v + \frac{\pa\Lc}{\pa\doa}\doa -
\frac{\dot{\pa\Lc}}{\pa\doa}a\right) - \Lc
\ee
for $\Lc(\tau; x,v,\dov,\ddov)$\,.
We see that for spinless NS's we have eq.\,(\ref{eq:Ham0}), that is the usual
Hamiltonian $pv-\Lc$.
The Hamiltonians involved by Lagrangians (\ref{eq:Lcn}) write:
\bb
\fbox{${\dis \Hc \ug \um\,mv^2}$}
\ee
for $\Lczero$;
\bb
\fbox{${\dis \Hc \ug \um\,m\,v^2 + \left(\um\,k_1\,a^2 -
k_1\,\doa v\right)}$}                                      \label{eq:Ham1}
\ee
for $\Lcuno$;
\bb
\fbox{${\dis \Hc \ug \um\,m\,v^2 + \left(\um\,k_1\,a^2 -
k_1\,\doa v\right) + \left(\um\,k_2\,\doa^2 + k_2\,\dddoa v
- k_2\,\ddoa a\right)}$}
\ee
for $\Lcdue$.

\

\

\noindent Let us now pass to write, for the first-order Lagrangians
$\Lc(\tau; x,v,\dov)$, ``Hamilton equations'' fully equivalent to the
Eulero-Lagrange equation.

Besides the first-order momentum
$p\equiv\pa\Lc/\pa\dox-\rd(\pa\Lc/\pa\ddox)/\rd\tau$
given by (\ref{eq:Mom}), let us define a ``second-order momentum'':
\bb
\pi \equiv \frac{\pa\Lc}{\pa\dov}\,.                       \label{eq:MomII}
\ee
Consequently the above Hamiltonian, eq.\,(\ref{eq:Ham01}), may be re-written as follows:
\bb
\Hc(\tau;\;x,p;\;v,\pi) \ug p\dox + \pi\dov - \Lc\,.
                                                           \label{eq:Ham1Dirac}
\ee
Using the differential $\rd\Hc$ of the Hamiltonian
$$
\rd\Hc = \dox\rd p+p\rd v+\dov\rd\pi+\pi\rd\dov-\rd\Lc\,,
$$
the above definitions of the momenta, and eq.\,(\ref{eq:dop}), we obtain
$$
\frac{\pa\Hc}{\pa x}=-\,\frac{\pa\Lc}{\pa x}=-\,\dop\,,
$$
$$
\frac{\pa\Hc}{\pa v}=p-\,\frac{\pa\Lc}{\pa v}=-\,\dot{\frac{\pa\Lc}{\pa\dov}}=-\,\dopi
$$
Thus we finally can write the following {\em double couple of Hamilton equations}

\

$\left\{\begin{array}{l}
{\dis\frac{\pa\Hc}{\pa p} \ug \dox}\\
\ \\
{\dis\frac{\pa\Hc}{\pa x} \ug -\dop}
\end{array}\right.
\hfill\left\{\begin{array}{l}
{\dis\frac{\pa\Hc}{\pa\pi} \ug \dov}\\
\ \\
{\dis\frac{\pa\Hc}{\pa v} \ug -\dopi} \ \ .
\end{array}\right.
\hfill \ $

\

\noindent Thus, besides the standard couple of Hamilton equations, we have
a new non-newtonian couple of Hamilton equations applying to the second-order
pair of canonical variables $(v, \pi)$. An identical result has been found in
ref.\cite{Pavsic}, but employing a different Hamiltonian.

\section{Classical Dirac particles} 

Since we are employing the proper time, let us shortly recall the so-called
``proper time formulation'' of the Dirac theory\cite{Staunton}.
In this formalism we can re-write the Dirac equation \
$\ppo\,\psi\ug m\,\psi$ \ in the form of a Schr\"odinger eingenvalues-equation,
introducing the scalar ``proper Hamiltonian'' $\Ho\equiv\ppo$:
\bb
\Ho\,\psi = i\,\pa_\tau\,\psi = m\psi\,,                   \label{eq:Y}
\ee
quantity $\tau$ being, as before, the CMF time and $m$ representing the energy
eigenvalue in the CMF. The Heisenberg equation for the proper-time
derivative of a generic operator $\Go$ writes:
\bb
\widehat{\dot{G}} \ug i\,[\Ho,\;\Go]\,.
\ee
Applying such an equation to the spacetime coordinate $x^\nu$ we get
\bb
\widehat{\dot{x}}^\nu \equiv \vo^\nu = i\,[\Ho,\;x^\nu] =
i\,[\po_\mu\gm,\;x^\nu] = \gn\,.                           \label{eq:xop}
\ee
Hereby we might say that {\em the quantum equivalent of classical constraint}
(\ref{eq:H2}), \ $\pp = m$, \ is actually the Dirac equation itself
\ $\ppo\,\psi=m\,\psi$. Also the classical conservation equations (\ref{eq:B})
and (\ref{eq:C}) are recovered in the operatorial form:
\bb
\widehat{\dop}\,^\nu = i\,[\Ho,\;\po^\nu] =
i\,[\po_\mu\gm,\;\po^\nu] = 0\,;                           \label{eq:BB1}
\ee
\bb
\widehat{\doS}\,^\munu = i\,[\Ho,\;\So^\munu] =
i\,\left[\po_\rho\ga^\rho,\;\frac{i}{4}\,(\gm\gn - \gn\gm)\right] =
\po^\mu\gn - \po^\nu\gm\,.                                 \label{eq:BB2}
\ee
For the 4-acceleration operator we have:
\bb
\ao^\nu = i\,[\Ho,\;\vo^\nu] = i\,[\po_\mu\gm,\;\gn] = 4\,\So^{\nu\rho}\po_\rho\,.
                                                           \label{eq:J}
\ee
Eqs.(\ref{eq:xop}-\ref{eq:J}) define the Dirac operatorial algebra
(cf.\,\cite{Hestenes,Barut}).

\

\noindent Let us now prove that the {\rm 1}-st order Lagrangian
\bb
\Lcuno \ug \um\,m\,v^2 + \um k_1\,a^2
\ee
can describe a free Dirac particle provided that we assume
\bb
k_1 \ug -\,\frac{1}{4m}\,,
\ee
so that
\bb
\fbox{${\dis\Lc_{\rm Dirac} \ug \um\,m\,v^2 -
\frac{a^2}{8m}}$}\,.                                       \label{eq:LagrDirac}
\ee
In fact equations (\ref{eq:BB1}) and (\ref{eq:BB2}), expressing the conservation
of 4-momentum and total angular momentum, are carried by any $\Lcn$. The third
equation
\bb
a^\mu\ug4\,S^\munu p_\nu                                   \label{eq:Z}
\ee
holds instead only in the special case of the Dirac theory, and is
substancially equivalent to the Eulero-Lagrange equation for $\Lcuno$
---i.e., to the generalized Newton equation (\ref{eq:NT})--- with
$k_1=-1/4m; \ k_i=0, i\geq 2$:
\bb
m\,a^\mu + \frac{1}{4m}\,\ddoa^\mu = 0\,.                  \label{eq:ZZ}
\ee
In fact, by derivating side by side eq.\,(\ref{eq:Z}) we get
\bb
\doa^\mu = 4\,\doS^\munu\,p_\nu\,,                         \label{eq:CC}
\ee
and then
\bb
\doS^\munu\,p_\nu \ug \frac{\doa^\mu}{4}\,.                \label{eq:doS}
\ee
But if we insert this result in the general zitterbewegung equation
(\ref{eq:I}) we obtain
\bb
m\,v^\mu + \frac{\ddov^\mu}{4m} \ug p^\mu\,,
                                                           \label{eq:AA}
\ee
which is nothing but eq.\,(\ref{eq:ZZ}) after its integration with respect to $\tau$.
Therefore the spin vector of a classical Dirac particle is given by
eq.\,(\ref{eq:spin1}) with $k_1=-1/4m$:
\bb
\sbf={1\over 4m}\,(\abf\times\vbf)\,;                      \label{eq:spindirac}
\ee
while the spin tensor (\ref{eq:Smunu}) now writes
\bb
S^\munu={1\over 4m}\,\left(a^\mu v^\nu - a^\nu v^\mu\right)\,.  \label{eq:spintensor}
\ee

\

\

\noindent The equation of the motion (\ref{eq:AA}) is a 4-vectorial,
constant-coefficients, second-order differential equation. Its
general solution writes
\bb
\fbox{${\dis v^\mu = \frac{p^\mu}{m} + E^\mu\,\cos(2m\tau) +
H^\mu\,\sin(2m\tau)}$}\,,
                                                           \label{eq:HH}
\ee
where, being equation (\ref{eq:AA}) of the second order, we can fix two
initial conditions by choosing the constant 4-vectors $E^\mu$ and $H^\mu$.
The general solution exhibits the special zitterbewegung foreseen by
Schr\"odinger\cite{Schroedinger} since we have an oscillating motion around the CM
with the characteristic ``Compton frequency'' $2m$. This result has been found
also in other, alternative approaches and in very particular models\cite{Hestenes,
Barut, Corben, Papapetrou, Salesi, Pavsic}.
The equivalence between the above oscillatory solution and the solutions
obtained in the quoted papers is often merely formal. In ref.\,\cite{Pavsic},
the time parameter $\tau$ is chosen, like in the present work, equal to the
CMF time. Nevertheless, because of the presence of square roots of $\dox^2$ in
the adopted action
[$H^\mu\equiv(\sqrt{\dox^2})^{-1}\,\rd/\dtau(\dox^\mu/\sqrt{\dox^2})$]
\bb
\Lc\equiv \sqrt{\dox^2}\,\left(m - \frac{\mu}{\sqrt{\dox^2}}\,H^2\right)\,,
                                                            \label{eq:LagrPav}
\ee
either lightlike or spacelike motions ($\dox^2\leq 0$) are forbidden if
we will finite and real \mbox{4-velocities}. By contrast, our theory allows these
motions: for example, we shall see later that the lightlike motion, $\dox^2=0$,
is the only involving both spin 1/2 and uniform motion in the CMF.
Moreover, we easily see, after explicitation, that Lagrangian (\ref{eq:LagrPav})
is {\em not} time-reversal invariant, differently from our $T$-symmetric
Lagrangian (\ref{eq:LagrDirac}). Consequently, Lagrangian (\ref{eq:LagrPav})
allows also unphysical non-stationary motions in the CMF, exponentially growing
or damping.

\

\noindent In the CMF, according to eq.\,(\ref{eq:spindirac}), the spin vector
is
\bb
\sbf\cmf\ug\um\,(\Hbf\cmf\times\Ebf\cmf)\,,                  \label{eq:FF}
\ee
and then it is orthogonal to the orbital plane defined by $\Ebf\cmf$ and
$\Hbf\cmf$. As expected, in the CMF the spin vector is time-constant, whilst in
a generic frame it is constant only its projection $\sbf_p$ along the
momentum (helicity), in its turn equal to the projection of $\sbf\cmf$ along
the momentum
$$
\sbf_p = (\sbf\cmf)_p =
\um\,(\Hbf\cmf^\perp\times\Ebf\cmf^\perp) = {\rm constant}\,,
$$
where $\Hbf\cmf^\perp$ and $\Ebf\cmf^\perp$ are, respectively, the components
of $\Hbf\cmf$ and $\Ebf\cmf$ orthogonal to $\imp$.

Because of the ``Correspondence Principle'' we can relate the (classical) spin
vector $\sbf\cmf$ in the CMF to the mean {\em non-relativistic} (quantum) spin
vector, averaged in a given (spin-$\um$) state
$\psi$, \ \mbox{$\ovsbf_{\rm qu}=\int\,\psid\sigbf\psi\,\rd V/2$.} \
We ask these vectors to have equal magnitudo. It is well-known\cite{Landau-3}
that the modulus of the mean spin vector, anyever be the considered
quantum state $\psi$, is always equal to 1/2:
\bb
|\ovsbf_{\rm qu}| = \um\,\left|\int\,\psid\sigbf\psi\rd V\right| = \um\,.
\ee
Taking account of eq.\,(\ref{eq:FF}) let us require
\bb
|\sbf\cmf| = \um\,|\Hbf\cmf\times\Ebf\cmf|
= \um\,.                                                   \label{eq:GG}
\ee
Hereby, by exploiting the known algebraic identity
\ $(\abf \times \bbf)^2 = \abf^2\bbf^2 - (\abf\cdot\bbf)^2$, \ we have the
following condition for Dirac spin-$\um$ particles:
\bb
\Ebf^2\cmf\Hbf^2\cmf - (\Ebf\cmf\cdot\Hbf\cmf)^2 = 1\,.    \label{eq:spin1/2}
\ee
[The same result can be obtained requiring that, in an arbitrary frame, the
(classical) value of $|\sbf|$ averaged over a zitterbewegung period be equal to the
magnitudo of the (quantized) helicity ($\um$)].
Notice that the above constraint, implying $\Ebf\cmf$ and $\Hbf\cmf$ to be not
parallel, does not allow pure linear oscillations for Dirac spin-$\um$
particles.

\

\noindent Besides the basic condition (\ref{eq:spin1/2}), from (\ref{eq:HH})
and from (\ref{eq:H2}) [or from (\ref{eq:orto})] we derive the following
useful constraint:
\bb
p_\mu E^\mu \ug p_\mu H^\mu \ug 0\,,                       \label{eq:pEpH}
\ee
that implies either $E^2, H^2 < 0$ (spacelike vectors), or $E^\mu=H^\mu=
(0;\,0,0,0)$ (null vectors). As we shall see below, the latter case implies a
vanishing spin [cf.\,(\ref{eq:FF})] and then refers only to spinless NS's, and
not to Dirac systems for which therefore the spacelike case always holds. In
the CMF, according to (\ref{eq:V0}) and to (\ref{eq:pEpH}), we always have
\bb
E^0\cmf \ug H^0\cmf \ug 0\,,                               \label{eq:MM}
\ee
that is, the CMF is a standard frame for the spacelike 4-vectors $E^\mu$ and
$H^\mu$.
From (\ref{eq:pEpH}) we have also
\bb
E^0=\wbf\cdot\Ebf\,,                                       \label{eq:E0}
\ee
and
\bb
H^0=\wbf\cdot\Hbf\,.                                       \label{eq:H0}
\ee
Let us write down the explicit expression of the 3-velocity referred to a generic
frame
\bb
\frac{\rd\xbf}{\dt}=\frac{\rd\xbf}{\dtau}\,\frac{\dtau}{\dt} \equiv
\frac{\vbf}{v^0} =
\frac{\imp+m\,\Ebf\,\cos(2m\tau)+
m\,\Hbf\,\sin(2m\tau)}{p^0+m\,\wbf\cdot\Ebf\,\cos(2m\tau)+
m\,\wbf\cdot\Hbf\,\sin(2m\tau)}\,,
\ee
where eqs.\,(\ref{eq:HH}), (\ref{eq:E0}) and (\ref{eq:H0}) have been applied.
In the absence of spin we have obviously the usual expression
$\rd\xbf/\rd t=\imp/p^0=\wbf$. The times-ratio
\bb
\frac{\dt}{\dtau}\equiv v^0=\frac{p^0}{m}+
\wbf\cdot\Ebf\,\cos(2m\tau) + \wbf\cdot\Hbf\,\sin(2m\tau)\,,
\ee
is time-oscillating around its mean value, namely around the Lorentz factor
$p^0/m=\gamma$. In general, during a zitterbewegung cycle we may observe
both time-dilation ($\dt/\dtau>1$) and time-contraction ($\dt/\dtau<1$), and
even time-inversion ($\dt/\dtau<0$). The overall effect, measured at the end of
each oscillation is, of course, the usual time-dilation: the ratio between two
zitterbewegung periods (referred to the laboratory and to the CMF, respectively)
is always equal to the Lorentz factor. An analogous non-constant relation holds
between the times elapsed in two generic reference systems:
\bb
\frac{\dt}{\dt^\prime}\equiv\frac{\dt}{\dtau}\frac{\dtau}{\dt^\prime}\equiv
\frac{v_0}{{v_0}^\prime} = \frac
{p_0 + m\,\wbf\cdot\Ebf\,\cos(2m\tau) + m\,\wbf\cdot\Hbf\,\sin(2m\tau)}
{p_0^\prime + m\,\wbf^\prime\cdot\Ebf\,\cos(2m\tau) +
m\,\wbf^\prime\cdot\Hbf\,\sin(2m\tau)}\,.
\ee

\

\noindent By integrating equation (\ref{eq:HH}) we derive the generic equation
of the trajectory of a free Dirac CS
\bb
x^\mu = x^\mu(0) + \frac{p^\mu}{m}\tau +
\frac{1}{2m}\,E^\mu\,\sin(2m\tau) - \frac{1}{2m}\,H^\mu\,\cos(2m\tau)\,.
                                                           \label{eq:NN}
\ee
Thus a Dirac point-like charge moves along a cylindrical helix, spiralizing
around the direction of the constant momentum. As said in subsection 1.3.3,
this happens not only in the ordinary 3-space, but also in spacetime since
the proper time $\tau$ and the laboratory time $t$ are not linearly linked as
for NS's and the times-ratio $v^0\equiv\rd t/\rd\tau$ oscillates.

Notice that the trajectory is a {\em right} helix if the plane of the ellyptical
orbit containing $\Ebf$ and $\Hbf$ is orthogonal to $\imp$: when
$(\Hbf\times\Ebf)\times\imp=0$ and $\imp\cdot\Vbf=0$.
In this case we shall have no longitudinal zitterbewegung along the straight
path of the CM. For what seen in subsection 1.3.2, this happens in the standard
frames for the 4-vector $V^\mu$, where $V^0(\!=\!V^0\cmf)=0$ and the times-ratio
$v^0=w^0+V^0$ is time-constant. Since $\imp$ is orthogonal to the zitterbewegung
plane containing $\Ebf(\!=\!\Ebf\cmf)$ and $\Hbf(\!=\!\Hbf\cmf)$ which
for (\ref{eq:FF}) is orthogonal to the spin, then $\imp$ results parallel
to $\sbf(\!=\!\sbf\cmf)$: i.e., {\em in the standard frames for $V^\mu$ CS's appear to be
polarized}. Summarizing: a polarized Dirac classical charge travels along a
right helix, without forwards and backwards oscillations along the direction
of the momentum, and the ratio between the CMF time and the laboratory time is the
usual constant Lorentz factor.

\

\

\noindent Let us now look for all the solutions of (\ref{eq:HH}) endowed with
{\em constant} $v^2$. By superimposing such a claim we obtain the following
conditions:
\bb
\left\{\begin{array}{l}
E_\mu H^\mu \ug 0\\

E^2 \ug H^2 \ .                                            \label{eq:orthnorm}\\
\end{array}\right.
\ee
Of course, the constraint $v^2$=constant does not involve in a generic frame a
uniform circular motion, but only ${v_0}^2-\vbf^2$=constant. In the CMF,
where $v_0=1$, we have instead a uniform circular motion, as it can be derived
also by inserting $p^\mu = (m; \ 0, \ 0, \ 0)$ and (\ref{eq:orthnorm}) in
(\ref{eq:HH}) or (\ref{eq:NN}). The orbital speed and the orbital radius are
\bb
|\vbf\cmf|\ug\sqrt{-E^2}\ug\sqrt{\Ebf\cmf^2}\,,
\ee
\bb
R\cmf = \frac{|\vbf\cmf|}{\omega} = \frac{|\vbf\cmf|}{2m} =
\frac{\sqrt{-E^2}}{2m} = \frac{\sqrt{\Ebf\cmf^2}}{2m}\,,   \label{eq:OO}
\ee
where the vectors $E$ may be indifferently replaced by the vectors $H$, since
from (\ref{eq:orthnorm}) we have $\Ebf\cmf^2=\Hbf\cmf^2$.
Always confining ourselves to the $v^2$-constant solutions, we easily derive from
(\ref{eq:spin1/2}) and (\ref{eq:orthnorm}) that $|\sbf|=1/2$ only for the
{\em lightlike} motion $v^2=0$,
$|\vbf\cmf|=1$, with $R\cmf$ equal to the Compton wavelenght $(2m)^{-1}$.
By inserting constraints (\ref{eq:orthnorm}) in eq.\,(\ref{eq:HH}), we get the
constant 4-velocity squared which results always less than 1 as expected from
(\ref{eq:v2<1}):
\bb
v^2 \ug 1 + E^2 \ug 1 + H^2 < 1\,.                         \label{eq:JJ}
\ee
Derivating now side by side eq.\,(\ref{eq:HH}) we get the 4-acceleration:
$$
a^\mu \ug -2m\,E^\mu\,\sin(2m\tau) + 2m\,H^\mu\,\cos(2m\tau)\,,
$$
from which
\bb
a^2 \ug 4\,m^2\,E^2 \ug 4\,m^2\,(v^2-1) < 0\,.
\ee
Notice that the 4-acceleration is always a spacelike 4-vector, any be
$v^2$ (even not constant), as expected from the general condition
$p_\mu a^\mu=0$.

\

\

\noindent The conserved Pauli-Lubanski ``spin 4-vector'' (whose square,
a Casimir-invariant of the Poincar\'e group, is equal to $-m^2\sbf\cmf^2$) is
defined as
\bb
W^\mu \equiv \um\,\vare^{\mu\nu\rho\sigma}J_{\nu\rho}p_\sigma
= \um\,\vare^{\mu\nu\rho\sigma}S_{\nu\rho}p_\sigma
= (\sbf\cdot\imp; \; p^0\sbf-\imp\times\kbf)               \label{eq:Wmu}
\ee
where the 3-vector $\kbf$ is the Lorentz-boosts generator $\kbf\equiv(S^{01},
S^{02}, S^{03})$. Here, for a classical Dirac particle, we have after some
algebra
\bb
W^0 \ug \um\,\imp\cdot(\Hbf\times\Ebf)\,,
\ee
\bb
\Wbf=\um\,\left[p^0\,(\Hbf\times\Ebf) + H^0\,(\Ebf\times\imp) +
E^0\,(\imp\times\Hbf)\right]\,.
\ee
Obviously in the CMF $\Wbf$ reduces to $m\sbf\cmf$ [cf. eq.\,(\ref{eq:FF})].

\

\noindent Another interesting quantity is the (non-constant) 3-vector
$\kbf$, averaged over a zitterbewegung cycle. In the CMF this vector
times $e/m$ is a sort of ``intrinsic electric-dipole momentum'', by contrast
with the intrinsic magnetic-dipole momentum, i.e., as usual, $\sbf$ times
$e/m$. In fact in the CMF $e\kbf\cmf/m$ is equal to the
intrinsic electric-dipole momentum $\dbf\cmf\equiv e\,\rbf\cmf$ (oscillating
{\em with zero average}). Actually, for eq.\,(\ref{eq:spintensor}) we have
$$
\kbf\cmf=\frac{{a^0}\cmf \vbf\cmf - {v^0}\cmf \abf\cmf}{4m}
= -\frac{1}{4m}\abf\cmf\,;
$$
finally, exploiting the harmonic relation between $\rbf\cmf$ and $\abf\cmf$, \
$\abf\cmf+4m^2\rbf\cmf=0$ \ [cf. eq.\,(\ref{eq:NN}) in the CMF with $\xbf(0)=0$],
\bb
\dbf\cmf \equiv e\,\rbf\cmf = -\,\frac{e}{4m^2}\,\abf\cmf =
\,\frac{e}{m}\,\kbf\cmf\,.
\ee
The proportionality factor $e/m$ might perhaps be defined as ``gyroelectric
factor''. Let us pass to a generic reference system other than the CMF.
A straight calculation employing (\ref{eq:HH}) gives, after averaging over
a period $T=2\pi/\omega=\pi/m$, a {\em non-zero} mean value for $\kbf$
\bb
\okbf = \,\frac{1}{2}\,(H^0\Ebf - E^0\Hbf)\,.              \label{eq:okbf}
\ee
From (\ref{eq:okbf}) we see that the above quantity is zero only if $E^0$ and
$H^0$ vanish\footnote{We have already seen that $\Ebf$ and $\Hbf$ cannot be
parallel for Dirac particles.}: i.e., for what above-seen, in the CMF and in all
the frames where the system appears polarized. In general we can say that for
a free classical Dirac particle $\okbf$ is non-zero depending on the
orientation of the spin.

\

\

\noindent At last let us re-formulate the classical theory of Dirac particles
in the hamiltonian formalism introduced in section 2.3.

The classical Dirac Hamiltonian writes, according to (\ref{eq:Ham1Dirac}) and
(\ref{eq:LagrDirac})
\bb
\Hc(\tau; \; x,p; \; v,\pi) \ug p\dox + \pi\dov - \Lc =
pv - 2\,m\pi^2 - \um\,mv^2\,,
\ee
where the momenta $p$ and $\pi$ are, according to (\ref{eq:Mom}) and
(\ref{eq:MomII}),
\bb
p \equiv \frac{\pa\Lc}{\pa\dox} - \dot{\frac{\pa\Lc}{\pa\dov}} =
mv + \frac{\ddov}{4m}\,,
\ee
\bb
\pi \equiv \frac{\pa\Lc}{\pa\dov} = -\,\frac{\dov}{4m}\,.
\ee
Finally, the Hamilton equations

\

$\left\{\begin{array}{l}
{\dis\frac{\pa\Hc}{\pa p} \ug \dox}\\
\ \\
{\dis\frac{\pa\Hc}{\pa x} \ug -\dop}
\end{array}\right.
\hfill\left\{\begin{array}{l}
{\dis\frac{\pa\Hc}{\pa\pi} \ug \dov}\\
\ \\
{\dis\frac{\pa\Hc}{\pa v} \ug -\dopi} \ \ ,
\end{array}\right.
\hfill \ $

\

\noindent now become

\

$\left\{\begin{array}{l}
{\dis v \ug \dox}\\
\ \\
{\dis 0 \ug -\,\dop}
\end{array}\right.
\hfill\left\{\begin{array}{l}
{\dis -\,4\,m\pi \ug \dov}\\
\ \\
{\dis p - mv \ug -\,\dopi} \ \ .
\end{array}\right.
\hfill \ $

\

\

\noindent By inserting the last but one equation in the last one we get out, as
expected, the Dirac equation of the motion, eq.\,(\ref{eq:AA})
$$
v = \frac{p}{m} + \frac{\ddov}{4m^2}\,.
$$

\section{Spinning systems with zero intrinsic angular momentum} 

Let us show that, for each $\Lcn$ with $n$ from 1 to $\infty$, can exist CS's
{\em endowed, in an arbitrary frame, with non-zero spin {\rm 3}-vector and zitterbewegung, but
with zero intrinsic} (i.e.: in the CMF) {\em angular momentum}:
\bb
\sbf\neq 0 \qquad\qquad\qquad \sbf\cmf=0\,.                \label{eq:spinzero}
\ee
The CS's satysfying eq.\,(\ref{eq:spinzero}) seem to be endowed ---besides
the usual orbital angular momentum $\lbf\equiv\xbf\times\imp$--- with a kind of
``extrinsic'' spin, which arises only in the presence of the ``external'' motion
the CM, and disappears in the CMF, just like the orbital angular momentum does.
It is sufficient to consider, for a chosen $\Lcn$, those solutions of the
Eulero-Lagrange equation which entail a {\em rectilinear oscillatory motion} in
the CMF.\footnote{The linear oscillatory motion is harmonic only for
$\Lcuno$; it is in general anharmonic for $n\geq 2$; notice that for $n\geq 2$
we might have a vanishing $\sbf\cmf$ without necessarily imposing a rectilinear
trajectory: cf. eqs.\,(\ref{eq:spin2}), (\ref{eq:spin3}).} To make an example,
for $\Lcuno$ ---which describes, as aforeseen, {\em also} Dirac spin-$\um$
particles--- it is enough to assume, whichever is the chosen value of $k_1$, \
$\Ebf\cmf\neq 0; \ \Hbf\cmf=0$, or $\Ebf\cmf=0; \ \Hbf\cmf\neq 0$, or
$\Ebf\cmf$ parallel to $\Hbf\cmf$, for obtaining a
linear harmonic motion in the CMF. As it easy to check from
eqs.\,(\ref{eq:spin1}-\ref{eq:spin3}) and from the analogous formulae for
$n>3$, in correspondence to CMF rectilinear motions the intrinsic angular
momentum $\sbf\cmf$ actually vanishes at any time since $\vbf\cmf$ and
its time derivatives are collinear vectors: \ $\vbf\cmf /\!/
\abf\cmf/\!/ \doabf\cmf /\!/ \ \cdots \ $. By contrast, in a frame
other than the CMF the space part of the spin tensor, $S^{ik}$, is in general
non-zero. Let us apply to the CMF an arbitrary boost $\wbf$.
Labelling with $\parallel$ ($\perp$) the components parallel (orthogonal) to
the boost, and taking into account that $v^0\cmf=1$ and $a^0\cmf=0$, we can write
for the Lorentz-transformed components of the 4-vectors $v^\mu$ and $a^\mu$
$$
v\para = \gamma^{-1}v\para\cmf + v^0\cmf w = \gamma^{-1}v\para\cmf + w
\qquad \qquad v\orto \ug v\orto\cmf\,,
$$
$$
a\para = \gamma^{-1}a\para\cmf + a^0\cmf w = \gamma^{-1}a\para\cmf
\qquad \qquad a\orto \ug a\orto\cmf\,,
$$
and so on for the higher-order derivatives of the velocity. As a consequence,
in the new frame the Lorentz-transformed velocity $\vbf$ and its
Lorentz-transformed derivatives {\em are not anymore collinear}, so that
$S^{ik}\neq 0$ and $\sbf\equiv(S^{23}, S^{31}, S^{12})\neq 0$.
It is easy to check that $\vbf$, $\abf$, etc. belong to the plane $\alpha$
containing $\vbf\cmf$, $\abf\cmf$, etc. and $\wbf$. Being for
eq.\,(\ref{eq:spin1})-(\ref{eq:spin3}) orthogonal to the plane $\alpha$,
the spin vector $\sbf$ is then normal to the momentum
$\imp\,(\equiv m\wbf/\sqrt{1-w^2})$. It follows that the helicity always
vanishes:
\bb
\lambda \equiv \frac{\sbf\cdot\imp}{|\sbf||\imp|} \ug 0\,. \label{eq:lambda}
\ee
This could be alternatively got from the well-known relativistic property
that in any reference frame the (time-constant) projection of the spin onto
the direction of the momentum is equal to the projection of $\sbf\cmf$ along
the same direction. Since $\sbf\cmf=0$ for (\ref{eq:spinzero}),
eq.\,(\ref{eq:lambda}) follows.

The Pauli-Lubanski 4-vector is here always a null vector:
\bb
W^\mu \ug (0; 0,0,0)\,.                                    \label{eq:W=0}
\ee
Then we have, from (\ref{eq:Wmu}) and (\ref{eq:W=0}) $\sbf\cdot\imp=0$,
$\sbf=\wbf\times\kbf$, so that, as expected, the helicity is 0 in any frame and
the spin 3-vector vanishes in the CMF where $\wbf=0$.

Let us choose a specific Lagrangian for a detailed picture of this phenomenon,
i.e., $\Lcuno$. By integrating the generalized Newton equation for $\Lcuno$,
that is eq.\,(\ref{eq:NT}) where $k_i=0$ for $i\geq 2$, we may get, by imposing
suitable boundary conditions, an oscillating {\em linear} motion. This motion
implies, as aforesaid, the vanishing of $\sbf\cmf$. We have:
\bb
v^\mu = \frac{p^\mu}{m} + F^\mu\,\cos(\omega\tau)\,,
\ee
where $\omega\equiv\sqrt{-m/k_1}$ ($k_1<0$).
Thus the space trajectory turns out to be a tilted
sinusoid-like path belonging to the aforesaid plane $\alpha$, with the nodal
axis parallel to $\imp$. Actually the trajectory is quite different from
the one of a spinless NS which is a straight line.

The spin vector is given by $k_1\,(\vbf\times\abf)$ for
eq.\,(\ref{eq:spin1}). Then we have, after some algebra (notice that, since the
Lorentz boost $\wbf$ does not affect the components orthogonal to $\imp$, we
have $\Fbf\times\imp=\Fbf^\perp\times\imp=\Fbf^\perp\cmf\times\imp$)
\bb
\fbox{${\dis \sbf\ug\frac{1}{\sqrt{\omega}}\,(\Fbf\cmf^\perp\times\imp)\,
\sin(\omega\tau)}$}\,.
\ee
The spin does not preceed anymore as the spin of a Dirac particle does, but
{\em linearly vibrates along a direction orthogonal to the momentum}.
Furthermore, the average over a zitterbewegung period of the above vector turns
out to be zero.
By contrast, the time average of the spin squared does not vanish and results
proportional to the square of the momentum:
\bb
\overline{\sbf^2}=
\frac{1}{2\omega}\,|\Fbf\cmf^\perp\times\imp|^2 =
\frac{1}{2\omega}\,{\Fbf\cmf^\perp}^2\imp^2\,.             \label{eq:spsq}
\ee
A {\em quantum} analogue of the present CS will be a particle
endowed with zero helicity and, at the same time, endowed with spin in
an arbitrary frame (different from the CMF). Such an object can be
found in a recent (quantum) theory by Ahluwalia and Kirchbach\cite{Ahluwalia}.
As is known, the usual spin-1 Proca equation ---due to the
transverse Lorenz 
constraint $\pa_\mu\psi^\mu=0$ which violates the
completeness relation--- does {\em not} describe the complete physical
content of the (1/2,1/2) representation space of the Lorentz proper group.
In the mentioned paper those authors write a (vector) wave-equation which
includes the Proca theory as a particular case and describes the whole
representation space. They also show that the (1/2,1/2) representations can be
divided into a triplet and a singlet of opposite relative intrinsic parities,
but {\em do not carry a definite spin angular momentum}. In general both
spin-1 and spin-0 particles are covariantly inseparable inhabitants of massive
vector fields.\footnote{From various considerations Ahluwalia and Kirchbach
conclude that the (1/2,1/2) space appears to be very suitable for a consistent
picture not only of W$^{\pm}$ and Z$^0$ vector gauge bosons but also
of Higgs bosons.} In particular, the state labelled\cite{Ahluwalia} as
$w_4$ does describe a particle endowed with spin 0 in the CMF, but {\em not} in
an arbitrary frame where we have a superposition of the helicity-0 $s=0$
and $s=1$ eigenstates. As a consequence that solution just refers to a
helicity-0 spinning particle.

\

\section{Summary of the results} 

The theory outlined in this paper appears very general because it is
not based on particular models or special approaches. The classical spin is
studied simply by generalizing and extending (through the usual tensorial
algebra) the newtonian theory, without any recourse to Grassmann variables or
special Clifford algebras. The only necessary (for free classical systems)
assumptions are spacetime isotropy and homogeneity.
Thus, starting from the conservation of the linear and angular momenta,
$\dop^\mu=\doJ^\munu=0$, in the first section we have obtained a zitterbewegung
equation, in which appear besides the timelike newtonian term $p/m$ also
a spacelike zitterbewegung-term. Among the consequences of the zitterbewegung
we have:

\ a) even in the absence of forces, differently from the momentum, the velocity
is not required to be a constant quantity: the Principle of
Inertia is not a general law for the classical free motion;
or equivalently, the RF is not an {\em inertial} reference frame;

\ b) the square of the 4-velocity $v^2$ obeys non-ordinary constraints;

\ c) ``global'' superluminal motions are not forbidden, provided that the
energy-momentum, and any related signal or information, move along the
worldline of the CM, which travels always with a subluminal speed;

\ d) in general, the zitterbewegung motion of a CS has a component along the
momentum;

\ e) the ratio between the time durations measured in a generic frame and in
the CMF is not constant and differs from the Lorentz factor (as instead occurs
for NS's). In general we can say that a non-linear relation occurs between the
time durations measured in different reference frames.

Newtonian mechanics and usual relativistic kinematics are of course recovered
as a {\em particular case} of the present theory: namely the spinless case with
no zitterbewegung. We have also analyzed the strict analogy holding between the
classical zitterbewegung equation and the quantum Gordon decomposition of the
Dirac current.

\

\noindent In the second section we have performed the lagrangian formulation of
our theory through a direct generalization (satisfying the relativistic
covariance and the symmetry under spacetime inversions) of the newtonian
Lagrangian to Lagrangians containing time derivatives of the velocity. The
results are obtained without any particular choose of the coefficients appearing
in the theory and without any ad-hoc assumption. We have derived a
constant-coefficients differential equation of the motion (generalization of
the Newton law $a=F/m$) in which non-newtonian zitterbewegung terms appear.
Alternate signs are requested for the coefficients of the terms appearing in
the Lagrangian if we want only stationary solutions and finite oscillatory
motions. Through the N\"other theorem, by satisfying the rotational symmetry,
the classical spin can be defined employing only classical kinematical
quantities, without recourse to quantum quantities as the Planck constant
$\hbar$, or to Grassmann non-commuting numbers. Imposing the
reparametrization invariance the conserved Hamiltonian is also obtained.
For the important case of the first-order Lagrangian it can be written
a second couple of Hamilton equations for the canonically conjugate
variables $v$ and $\pi$, in addition to the usual couple of Hamilton equations
referring to the canonical variables $x$ and $p$.

\

\noindent In the third section we have shown that the first-order Lagrangian with
$k_1=-1/4m$ fully describes classical Dirac particles and derived the
classical Dirac spin in the form $\abf\times\vbf/4m$. The general solution of
the Eulero-Lagrange equation oscillates with the Compton frequency $\omega=2m$,
and the spacetime worldline is a 4-dimensional helix. The particular solutions
corresponding to polarized Dirac particles and to constant $v^2$ have been in
detail studied.
Finally, we have derived the explicit form of the Pauli-Lubanski 4-vector for
classical Dirac particles.

\

\noindent In the last section we have studied spinning CS's with zero
intrinsic angular momentum and helicity, ispecially in the case of first-order
Lagrangians. An interesting quantum analogue has been recently found
in a (1/2,1/2) vector representation of the Lorentz group, shown to be a
superposition of spin-0 and spin-1 states, with $s=0$ only \mbox{in the CMF.}

\

\

\

{\bf Acknowledgements}

\noindent The author is glad to thank E.\,Recami and the Referee for
helpful hints and for having suggested some bibliography. The scientific
collaboration of S.\,Esposito and D.\,Zappal\`a is also acknowledged. \ For the
kind, active cooperation thanks are due to G.\,Andronico, G.G.N. Angilella,
P.\,Falsaperla, G.\,Giuffrida, S.\,Lo Nigro, T.\,Venasco and B.\,Zappa.\\

\

\

\

\end{document}